\documentclass[prd,twocolumn,showpacs]{revtex4}
\usepackage{graphicx}
\usepackage{dcolumn}
\usepackage{bm}
\usepackage{natbib}
\usepackage{amssymb}

\newcommand{\bpsi}{{\boldsymbol\psi}}

\begin{document}

\title{Resonant recoil in extreme mass ratio binary black hole mergers}

\author{Christopher M. Hirata}
\affiliation{Caltech M/C 350-17, Pasadena CA 91125, USA}
\email{chirata@tapir.caltech.edu}

\date{June 8, 2011}

\begin{abstract}
The inspiral and merger of a binary black hole system generally leads to an asymmetric distribution of emitted radiation, and hence a recoil of the remnant black hole directed opposite to the net linear momentum radiated.  The recoil velocity is generally largest for comparable mass black holes and particular spin configurations, and approaches zero in the extreme mass ratio limit.  It is generally believed that for extreme mass ratios $\eta\ll 1$, the scaling of the recoil velocity is $|{\bf V}|\propto\eta^2$, where the proportionality coefficient depends on the spin of the larger hole and the geometry of the system (e.g. orbital inclination). The small recoil velocity is due to cancellations: while the fraction of the total binary mass radiated away in gravitational waves is $O(\eta)$, most of this energy is emitted during the inspiral phase where the momentum radiated integrates to zero over an orbit.  Here we show that for low but nonzero inclination prograde orbits and very rapidly spinning large holes (spin parameter $a_\star>0.9678$) the inspiralling binary can pass through resonances where the orbit-averaged radiation-reaction force is nonzero.  These resonance crossings lead to a new contribution to the kick, $|{\bf V}|\propto\eta^{3/2}$.  For these configurations and sufficiently extreme mass ratios, this resonant recoil is dominant.  While it seems doubtful that the resonant recoil will be astrophysically significant, its existence suggests caution when extrapolating the results of numerical kick results to extreme mass ratios and near-maximal spins.
\end{abstract}

\pacs{04.30.Db, 04.25.Nx, 04.70.Bw}

\maketitle

\section{Introduction}

It has long been recognized that gravitational waves from an asymmetric source can result in a net emission of linear momentum and a consequent recoil or ``kick'' of the system \cite{1973ApJ...183..657B}.  One of the most important realizations of this scenario is in the merger of a binary black hole system \cite{1983MNRAS.203.1049F}, which has attracted the interest of the astrophysics community due to the possibility of disturbing or ejecting massive black holes from the centers of galaxies and globular clusters \cite{1989ComAp..14..165R, 2002ApJ...581..438M, 2004ApJ...604..484M, 2004ApJ...606L..17M, 2004ApJ...607L...9M, 2004ApJ...613...36H, 2007ApJ...663L...5V, 2008AN....329.1004G, 2008ApJ...678..780G, 2008ApJ...686..829H}.

In recent years, several approaches have been used to compute binary black hole kicks.  For comparable masses and in the strong field part of the merger, numerical GR computations \cite{2005PhRvL..95l1101P, 2006ApJ...653L..93B, 2007ApJ...659L...5C, 2007CQGra..24...33H, 2007PhRvL..98i1101G, 2007PhRvL..98w1101G, 2007PhRvL..98w1102C, 2007ApJ...668.1140B, 2007PhRvD..76f1502T, 2007PhRvD..76h4032H, 2007PhRvD..76l4002P, 2008PhRvD..77l4047B, 2008ApJ...682L..29B, 2008PhRvD..78b4039D, 2009PhRvD..79f4018L, 2007PhRvL..99d1102K} now provide the best tool.  The post-Newtonian series is appropriate for computing the waveform emitted during inspiral, and at sufficiently high order can even follow the plunge phase \cite{2005ApJ...635..508B, 2010CQGra..27a2001L}.  For extreme mass ratio inspirals (EMRIs), one may use black hole perturbation theory (BHPT), which treats the smaller (``secondary'') black hole as a test particle in the metric of the primary hole, with slowly varying constants of motion to account for radiation reaction followed by geodesic motion during the plunge phase \cite{1984MNRAS.211..933F, 2004ApJ...607L...5F, 2008PhRvD..78l4015M, 2010PhRvD..81j4009S}.  One may also apply BHPT to the ringdown phase in the form of the ``close limit approximation,'' which is based on perturbations around the final black hole state \cite{1994PhRvL..72.3297P, 2006PhRvD..74l4010S}.  There have also been kick computations \cite{2006PhRvD..73l4006D, 2007ApJ...662L..63S} using the ``effective one-body'' (EOB) method, which maps the finite mass ratio problem to the motion of a test particle in a modified black hole metric whose form is constrained by agreement with the post-Newtonian expansion \cite{1999PhRvD..59h4006B}; and there are computations based on BHPT waveforms but with source terms determined from EOB inspiral trajectories \cite{2010PhRvD..81h4056B}.  Most of the analyses have focused on inspiral from initially circular orbits, since at large radii gravitational radiation tends to circularize the orbits on less than a merger timescale; but the same techniques have been applied to mergers with eccentric initial conditions \cite{2007ApJ...656L...9S}.

The computational expense of some of these approaches has motivated several researchers to propose kick ``fitting formulae'' that return the recoil velocity ${\bf V}$ of the remnant black hole as a function of initial masses, spins, and orbital parameters \cite{2007ApJ...659L...5C, 2007ApJ...668.1140B, 2008ApJ...682L..29B, 2010CQGra..27k4006L, 2010ApJ...719.1427V, 2010arXiv1011.0593L, 2010arXiv1011.2210Z}.  Ideally, such a formula would both reproduce numerical GR computations and approach BHPT results in the extreme mass ratio limit.  It should also respect rotation, reflection, and interchange symmetries \cite{2008PhRvL.100o1101B, 2008PhRvD..78b4017B}.

The extreme mass ratio limit of black hole merger kicks is perhaps less important from an astrophysical perspective because the kicks are small compared to the potential wells of galaxies.  Nevertheless, it is still of substantial theoretical interest: the (moderately) large mass ratio case is already starting to enable a comparison of perturbation theory to full numerical simulations in a regime where both are practical and valid (e.g. \cite{2009PhRvD..79l4006G, 2010PhRvD..81j4009S}); and BHPT arguments are used to set the limiting behavior of the kick fitting formulae, which are often used by astrophysicists.  Given that the fitting formulae are typically constrained with only a modest suite of simulations that sparsely samples parameter space, it is important to check their validity in any regime possible.

BHPT arguments typically give kicks that scale as $\propto \eta^2$ for EMRIs, where $\eta=\mu/M\ll 1$ is the mass ratio \cite{1984MNRAS.211..933F}.  (Since we work to lowest order in $\mu$, we will not distinguish here between $\eta$ and the traditional mass ratio $q=m_{\rm small}/m_{\rm large}$; and we will take $\mu$ and $M$ to be the smaller and larger masses, respectively.)  The reason is that even though the total energy emitted during the inspiral is $\sim\mu$, emitted in $\sim\eta^{-1}$ cycles, in problems considered thus far it is emitted symmetrically: averaged over an orbit, the amount of power radiated in direction $\hat{\bf n}$ is equal to that radiated in direction $-\hat{\bf n}$, resulting in zero net recoil.  The exception occurs as the inspiral ends and terminates in a plunge and finally a ringdown, which occurs over only of order one dynamical timescale ($\sim M$) and emits an amount of gravitational wave energy $\sim\mu^2/M$ in an asymmetric pattern.  This leads to a kick only of order $V\sim (\mu^2/M)/M \sim\eta^2$, rather than $\eta$, since only the energy radiated in the last dynamical time produces a unidirectional force. In the case of nonspinning black holes as $\eta\rightarrow 0$, recent perturbative computations have found $V = 0.0446\eta^2$ \cite{2010PhRvD..81h4056B} and $0.044\eta^2$ \cite{2010PhRvD..81j4009S}, in agreement with the estimate of $0.04396\eta^2$ \cite{2010PhRvD..81h4056B} obtained via extrapolation of numerical simulations \cite{2007PhRvL..98i1101G, 2009PhRvD..79l4006G}. The case of EMRIs in the equatorial plane of a spinning primary hole also leads to $V\propto\eta^2$, but with a coefficient that depends on the primary spin $a_\star$ \cite{2010PhRvD..81j4009S}.

The principal purpose of this paper is to argue that in some cases, an EMRI can produce a kick with a limiting behavior $V\sim\eta^{3/2}$ instead of $\sim\eta^2$.  The required conditions -- at least for circular inspirals, which are the sole focus of this paper -- are (i) a very large spin for the primary black hole, $a_\star>0.9678$; and (ii) a low but nonzero inclination prograde orbit.  Under such circumstances, the inspiralling binary can pass through resonances between the vertical and azimuthal frequencies: specifically, the increase in longitude between successive ascending node passages $\Phi$ can be an integer multiple of $4\pi$ (instead of having $\Phi=2\pi$ as occurs in any spherically symmetric spacetime, such as Schwarzschild).  When the system is in such a resonance, the orbit-averaged recoil force is {\em not} zero, but rather $\sim\eta^2$.  The EMRI cannot be trapped in this resonance since there is no preferred longitude in the problem, but rather it continues its inspiral; as a result, the resonant argument $\varphi$ switches its direction of circulation ($\dot\varphi$ changes sign).  During the resonance crossing, the resonant argument has roughly constant phase (varying by $\lesssim 1$ radian) for a duration of time $t_\varphi\sim |\ddot\varphi|^{-1/2}$; since $\dot\varphi$ is a combination of orbital frequencies (of order $M^{-1}$) that varies on the inspiral timescale $M^2/\mu$, we have $\ddot\varphi\sim \mu/M^3$ and hence $t_\varphi\sim \mu^{-1/2}M^{3/2}$.  Putting this together, we find that the resonant kick is the orbit-averaged force times the dephasing time, divided by the mass of the system: $V\sim (\eta^2)(\mu^{-1/2}M^{3/2})/M\sim \eta^{3/2}$.  In this paper, we will use the method of stationary phase to demonstrate this scaling, and explicitly evaluate the prefactor for some values of spin parameter $a_\star$ and inclination $\iota$.

The resonant kick dominates over the ${\cal O}(\eta^2)$ transition/plunge kick for sufficiently extreme mass ratios, in those cases where it occurs.  In fact, for most of the parameter space in $(a_\star,\iota)$ where a resonance crossing occurs it appears likely that the resonant kick will dominate.  We note that in cases where both kicks are comparable, the overall magnitude of the kick will depend very sensitively on initial conditions, because -- as the vector sum of two contributions -- it will depend on the relative longitude of the ascending node at the resonance crossing and the longitude at plunge, as well as the phase of the vertical oscillation at plunge.

None of the currently published kick velocity fitting formulae contain an order $\eta^{3/2}$ contribution.  While it is not clear whether the resonant kick (or its intermediate mass ratio analogue) is significant for astrophysical cases, the existence of the $\eta^{3/2}$ scaling in some part of $(a_\star,\iota)$ space suggests caution when constructing fitting functions or extrapolating numerical GR results to extreme mass ratios and/or spins.

We note that previous work on EMRIs has identified the radial-vertical resonances in generic (eccentric and inclined) inspirals as potentially important for waveform computation: they yield a deviation from adiabatic inspiral behavior at resonance crossings due to radiation reaction and self-force corrections \cite{2005PThPh.113..733M, 2010arXiv1009.4923F}, or in spacetimes that deviate from the Kerr solution \cite{2009PhRvL.103k1101A}.  However these resonances do not exist for circular orbits, and none of these analyses appear to have considered the effect on the radiated linear momentum.

This paper is organized as follows.  The theoretical arguments are presented in \S\ref{S:th}.  The computation and the resulting kick magnitudes are given in \S\ref{S:r}, and their significance is discussed in \S\ref{S:d}.  The formalism and associated code are described at length in Ref.~\cite{2010arXiv1010.0759H}; we repeat only the most important points here and refer the reader to Ref.~\cite{2010arXiv1010.0759H} for implementation details.  We use relativistic units where $G=c=1$.

\section{Theory}
\label{S:th}

We treat the motion of the smaller black hole in the Boyer-Lindquist coordinate system \cite{1967JMP.....8..265B}, in which the Hamilton-Jacobi equation for the particle motion \cite{1968PhRv..174.1559C} and the Teukolsky equation \cite{1972PhRvL..29.1114T, 1973ApJ...185..635T, 1973ApJ...185..649P, 1974ApJ...193..443T} for the Weyl tensor perturbations are separable.  The metric in this system is
\begin{eqnarray}
ds^2 &=& 
-\left( 1 - \frac{2Mr}\Sigma \right) d t^2 - \frac{4Mar}\Sigma\sin^2\theta\, d t\,d\phi
\nonumber \\ &&
 + \frac{(r^2+a^2)^2
-\Delta a^2\sin^2\theta}\Sigma\,\sin^2\theta\,d\phi^2
\nonumber \\ && + \frac\Sigma\Delta\,d r^2 + \Sigma\,d\theta^2,
\end{eqnarray}
where $\Delta \equiv r^2 - 2Mr + a^2$ and
$\Sigma \equiv r^2+a^2\cos^2\theta$.
This reduces to standard spherical Minkowski coordinates at large $r$.  The mass of the hole is $M$ and its specific angular momentum is $a<M$; we denote the spin as a fraction of the maximal value by $a_\star \equiv a/M$.

Greek indices $\alpha\beta...$ will run over all four coordinates $\{t,r,\theta,\phi\}$, Latin indices $ij...$ will run over only the spatial coordinates $\{r,\theta,\phi\}$, and the overdot will denote a derivative with respect to coordinate time, $\dot{}=d/dt$.

\subsection{Geodesic motion}

We treat the motion of a test particle in the Kerr spacetime using the 3+1 Hamiltonian formalism and utilize a canonical transformation from the original $(x^i,p_i)$ variables to action-angle variables, following the notation and methodology of Ref.~\cite{2010arXiv1010.0759H}.  Note that this results in a different set of angle variables than Refs.~\cite{2008PhRvD..78f4028H, 2010arXiv1009.4923F}, whose 4-dimensional analyses use the proper time or Mino time \cite{2003PhRvD..67h4027M, 2005CQGra..22S.801D} and promote $t$ to be a dynamical variable with a conjugate momentum $p_t$, although the actions $J_i$ are the same.

In this picture, the action is given by the usual test particle formula $S=\int L\,dt$, where the Lagrangian is given by $L=-\mu\,d\tau/dt$, where $d\tau$ is the proper time element.  The action is varied with respect to the trajectory $\{r(t),\theta(t),\phi(t)\}$.  The conjugate momenta $(p_r,p_\theta,p_\phi)$ are easily seen to equal the covariant momentum components, and the Hamiltonian $H = p_i \dot x^i - L$ is easily seen to equal $H=-p_t$, the covariant $t$-component of the momentum determined by the mass-shell condition $g^{\alpha\beta}p_\alpha p_\beta = -\mu^2$.

The Kerr problem admits three constants of the motion: the Hamiltonian per unit mass ${\cal E}=-p_t/\mu$, the angular momentum per unit mass ${\cal L}=p_\phi/\mu$, and the Carter constant ${\cal Q}$; and these mutually commute: $\{{\cal E},{\cal L}\}_{\rm P}={\{\cal E},{\cal Q}\}_{\rm P}=\{{\cal L},{\cal Q}\}_{\rm P}=0$, where $\{,\}_{\rm P}$ represents a Poisson bracket.  Therefore the particle moves on a 3-torus of constant $({\cal E},{\cal Q},{\cal L})$, which may alternatively be parameterized by the action variables:
\begin{equation}
J_i = \frac1{2\pi} \oint_{C_i} p_j\,dx^j,
\end{equation}
where $C_i$ is a loop on a surface of constant $({\cal E},{\cal Q},{\cal L})$ where $i\in\{r,\theta,\phi\}$ advances through one cycle.  One may also 
write the reduced actions $\tilde J_i=J_i/\mu$, which depend only on the trajectory and not on $\mu$.  We may define the Jacobian of the transformation 
between the two parameterizations of the tori, $M_{Ai} = \partial K_A/\partial \tilde J_i$ or $[{\bf M}^{-1}]_{iA} = \partial\tilde J_i/\partial K_A$, 
where $K_A$ is one of ${\cal E}$, ${\cal Q}$, or ${\cal L}$.  We note that one of the actions is simply the angular momentum: $\tilde J_\phi={\cal L}$.

We also need the angle coordinates on these tori, $0\le \psi^i<2\pi$; the mapping from $(J_j,\psi_j)\rightarrow(x^i,p_i)$ may be explicitly constructed using (i) the direct conditions
\begin{eqnarray}
\left.\frac{\partial x^i}{\partial \psi^j}\right|_{\bf J} \!\! &=& \!\! +\mu [{\bf M}^{-1}]_{jA}\left.\frac{\partial K_A}{\partial p_i}\right|_{\bf x}
{\rm ~~and}\nonumber \\
\left.\frac{\partial p_i}{\partial \psi^j}\right|_{\bf J} \!\! &=& \!\! -\mu [{\bf M}^{-1}]_{jA}\left.\frac{\partial K_A}{\partial x^i}\right|_{\bf p};
\end{eqnarray}
and (ii) an initial condition or choice of origin, i.e. a point on the torus to define $\psi^r=\psi^\theta=\psi^\phi=0$.  There is some freedom in choosing the origin, but it is not arbitrary (the canonical conditions for $J_i$ and $\psi^i$ impose some constraints); a valid choice is to set $\psi^i=0$ at pericentre ($r=r_{\rm min}$), ascending node ($\theta=\pi/2$, $\dot\theta<0$), and zero longitude ($\phi=0$).  

The Hamiltonian is a function only of the actions and so the equations of motion become trivial: ${\bf J}$ is constant and
the angles advance at a constant rate, $\dot \psi^i = M_{{\cal E}i} \equiv \Omega_i$; we may thus write the angle solution as
\begin{equation}
{\bpsi} = {\bpsi}^{(0)} + {\bf\Omega}t,
\end{equation}
where ${\bpsi}^{(0)}$ represents the three initial phases.

In the Keplerian limit, $r\gg M$, all three frequencies become equal, $\Omega_r\approx \Omega_\theta\approx \Omega_\phi$.  For circular but inclined orbits, of interest here, we generally have a precession rate (i.e. rate of increase of the longitude of the ascending node) $\Omega_\phi-\Omega_\theta\neq 0$.

\subsection{Emitted waveform, power, and momentum}

BHPT enables us to compute the waveform emitted by an orbiting particle in the Kerr spacetime to linear order in its mass $\mu$.  It is a practical method of computation for EMRIs since it is valid in the strong-field regime, with the mass ratio as the only expansion parameter.  Computation of the orbit-averaged energy and angular momentum flux is by this point a ``standard'' problem and allows one to compute the adiabatic evolution of either circular or equatorial orbits around Kerr black holes \cite{1978ApJ...225..687D, 1993PhRvD..48..663S, 1994PhRvD..50.6297S, 1998PhRvD..58f4012K, 2000PhRvD..61h4004H}.  For generic (eccentric and inclined) orbits in Kerr, one also needs to compute the rate of change of the Carter constant $\dot{\cal Q}$ \cite{2003PhRvD..67h4027M, 2005PhRvL..94v1101H, 2006PhRvD..73b4027D}, which is one of the more difficult problems tackled by BHPT; but since we restrict to circular orbits this will not be necessary here.
 
The gravitational wave signal is encoded by $\psi_4$, the perturbation of the Weyl tensor component (not to be confused with an angle variable).  It obeys a wave equation with a source ${\cal T}(t,r,\theta,\phi)$ \cite{1973ApJ...185..635T}.  This equation turns out to be separable in the 4 coordinates (the $t$ and $\phi$ dependences follow from symmetry arguments, but the separation of the $r$ and $\theta$ dependences is nontrivial), and it possesses 3 separation constants $\{m,\omega,\lambda\}$: those associated with the longitude and time dependences $m$ and $\omega$ ($\psi_4\propto e^{im\phi} e^{-i\omega t}$) and the eigenvalue $\lambda$ of the $\theta$ equation (which has boundaries at $\theta=0$ and $\pi$).  For geodesic motion, the source is quasiperiodic in the sense that in its Fourier transform ${\cal T}_m(\omega,r,\theta)$ contains only frequencies $\omega = {\bf q}\cdot{\bf\Omega}$, where ${\bf q}\in{\mathbb Z}^3$ is a lattice vector.  The retarded solution for the gravitational waveform is
\begin{eqnarray}
\psi_4(t,r,\theta,\phi) \!\!\! &=& \!\!\! (r-ia\cos\theta)^{-4} \sum_{\ell m{\bf q}} Z^{\rm out}_{\ell m{\bf q}} {\cal R}_{3;\ell m\omega}(r)
\nonumber \\ && \times
S_{\ell m \omega}(\theta) e^{i(m\phi- {\bf q}\cdot{\bpsi}^{(0)} -\omega t)}.
\label{eq:Out}
\end{eqnarray}
In this equation:
\begin{itemize}
\item
$S_{\ell m\omega}(\theta)$ is an angular eigenfunction normalized by $\int_0^\pi |S_{\ell m\omega}(\theta)|^2\sin\theta\,d\theta=1$.  It is real, and it reduces to a spin-weighted spherical harmonic \cite{1966JMP.....7..863N, 1967JMP.....8.2155G} at zero longitude, $S_{\ell m\omega}(\theta) = _{\,-2\,}\!Y_{\ell m}(\theta,\phi=0)$, in the limit of a Schwarzschild black hole.  The azimuthal quantum number $m\in{\mathbb Z}$, and the vertical quantum number $\ell$ is an integer with $\ell\ge\max\{|m|,2\}$.  
\item
${\cal R}_{3;\ell m\omega}(r)$ is the purely-outgoing solution to the radial Teukolsky equation \cite{1973ApJ...185..635T};
its large-$r$ behavior is
\begin{equation}
(r-ia\cos\theta)^{-4} {\cal R}_{3;\ell m\omega}(r) \rightarrow r^{-1} e^{i\omega r_\star}.
\end{equation}
Here $r_\star(r)$ is a tortoise coordinate whose large-$r$ behavior is $r_\star = r - 2M\ln r + $constant; this accounts for the logarithmically divergent phase shift of emitted radiation in a $1/r$ potential.
\item
$Z^{\rm out}_{\ell m{\bf q}}$ is the emitted wave amplitude associated with the $\ell m$ gravitational wave mode, emitted by the ${\bf q}$ Fourier component (i.e. frequency $\omega={\bf q}\cdot{\bf\Omega}$).
\end{itemize}

We will consider circular ($J_r=0$) orbits in our problem, in which case there is no dependence of the source ${\cal T}$ on the radial phase; hence we may consider only Fourier modes on the torus with $q_r=0$.  Also the longitude shift symmetry guarantees that the $m$ Fourier mode of the waveform is contributed only by torus Fourier modes with $q_\phi=m$.  We may thus write ${\bf q}=(q_r,q_\theta,q_\phi)=(0,k,m)$.  Using this, and the fact that the emitted waveform amplitude is proportional to the particle mass $\mu$, we may use the notation $Z^{\rm out}_{\ell m{\bf q}} = \mu\tilde Z^{\rm out}_{\ell mk}$, where $\tilde Z^{\rm out}_{\ell mk}$ is independent of $\mu$.

[Note that Ref.~\cite{2010arXiv1010.0759H} drops the ${\bpsi}^{(0)}$ term in Eq.~(\ref{eq:Out}) since there it amounts to an arbitrary definition of longitude, but in our problem we do not have this luxury.]

The power emitted to future null infinity per unit solid angle is
\begin{equation}
\frac{dE}{dt\,d^2\hat{\bf n}} = \frac1{4\pi} \left\langle \left| \int r\psi_4\,dt \right|^2 \right\rangle,
\end{equation}
where the average value is meant to be taken over many cycles.  Here $\int \psi_4\,dt$ can be obtained by inserting a factor of $i/\omega$ in Eq.~(\ref{eq:Out}):
\begin{equation}
\frac{dE}{dt\,d^2\hat{\bf n}} = \frac{\mu^2}{4\pi} \left\langle \left|
\sum'_{\ell mk} \frac{\tilde Z^{\rm out}_{\ell mk}}{\omega}
S_{\ell m \omega}(\theta) e^{i(m\phi- {\bf q}\cdot{\bpsi}^{(0)} -\omega t)}
\right|^2 \right\rangle.
\label{eq:dEdn}
\end{equation}
The $'$ in the summation means that we drop terms with $\omega=0$, as these do not correspond to any radiated power.
The emitted power and force are then
\begin{equation}
\dot E_{\rm em} = \int_{S^2} \frac{dE}{dt\,d^2\hat{\bf n}} \,d^2\hat{\bf n}
{\rm ~~and~~}
{\bf F} = \int_{S^2} \hat{\bf n}\,\frac{dE}{dt\,d^2\hat{\bf n}} \,d^2\hat{\bf n}.
\end{equation}
The back-reaction force on the inspiralling binary is $-{\bf F}$ in accordance with conservation of momentum.

\subsection{Resonances}

If $\Omega_\phi/\Omega_\theta$ is irrational, then the average value in Eq.~(\ref{eq:dEdn}) contains no interference between different $(m,k)$.  If however there is a rational ratio $\Omega_\phi/\Omega_\theta$, then different values of $(m,k)$ can have the same frequency.  We take the resonant relation
\begin{equation}
n_\theta\Omega_\theta + n_\phi\Omega_\phi = 0
\label{eq:rescond}
\end{equation}
with $n_\theta$ and $n_\phi$ relatively prime, so that $\Omega_\phi/\Omega_\theta = -n_\theta/n_\phi$.  
Then there will be interference between two terms $(m_1,k_1)$ and $(m_2,k_2)=(m_1+\Delta m,k_1+\Delta k)$ if $\Omega_\phi\Delta m+\Omega_\theta\Delta k=0$.  This condition is satisfied when and only when
\begin{equation}
(\Delta m,\Delta k) = (pn_\phi,pn_\theta)
\end{equation}
for some $p\in{\mathbb Z}$.  In this case the two modes have the same frequency,
\begin{equation}
\omega = m_1\Omega_\phi + k_1\Omega_\theta = m_2\Omega_\phi + k_2\Omega_\theta.
\end{equation}

We may now expand Eq.~(\ref{eq:dEdn}) in the resonant case as
\begin{eqnarray}
\frac{dE}{dt\,d^2\hat{\bf n}} \!\!\! &=& \!\!\!
\frac{\mu^2}{4\pi} 
\sum'_{\ell_1 m_1 k_1 \ell_2 p}
\omega^{-2} \tilde Z^{{\rm out}\ast}_{\ell_1 m_1k_1} \tilde Z^{{\rm out}}_{\ell_2 m_2k_2}
\nonumber \\ && \!\!\! \times
S_{\ell_1 m_1 \omega}(\theta) S_{\ell_2 m_2 \omega}(\theta)
\nonumber \\ && \!\!\!
\times \exp [ipn_\phi\phi - ipn_\phi \psi^{\phi(0)} - ipn_\theta \psi^{\theta(0)} ].
\end{eqnarray}
In this equation the $^{(0)}$ superscripts on $\psi^i$ are inconsequential because $n_\phi \Omega_\phi + n_\theta\Omega_\theta=0$ and hence the phase factor is unchanged if we replace $\psi^{i(0)}\rightarrow \psi^i$:
\begin{eqnarray}
\frac{dE}{dt\,d^2\hat{\bf n}} \!\!\! &=& \!\!\!
\frac{\mu^2}{4\pi}
\sum'_{\ell_1 m_1 k_1 \ell_2 p}
\omega^{-2}\tilde Z^{{\rm out}\ast}_{\ell_1 m_1k_1} \tilde Z^{{\rm out}}_{\ell_2 m_2k_2}
\nonumber \\ && \!\!\! \times
S_{\ell_1 m_1 \omega}(\theta) S_{\ell_2 m_2 \omega}(\theta)
\nonumber \\ && \!\!\!
\times \exp (ipn_\phi\phi - ipn_\phi \psi^\phi + ipn_\theta \psi^\theta ).
\label{eq:FluxEm}
\end{eqnarray}
This is the form of the emitted power that we will use.  It contains both the torus-averaged contribution to the emitted power ($p=0$) and the contributions associated with the resonant orbit only wrapping around part of the torus ($p\neq 0$).  Our principal interest will be in the emitted power and force; we see that:
\begin{itemize}
\item
The $S^2$-integrated power $\dot E_{\rm em}$ is not affected by the resonant ($p\neq 0$) terms because the longitude integral of $\exp(ipn_\phi\phi)$ vanishes.
\item
Similarly, the $z$-component of the force $F_z$ is not affected by the resonance.  Moreover, the symmetry of the torus across the equator then guarantees that there is no time-averaged $z$ component to the force in the inspiral phase.
\item
The in-plane ($x$ and $y$) components of the force can receive contributions only from terms with $pn_\phi=\pm 1$, i.e. terms in Eq.~(\ref{eq:FluxEm}) that have nonzero $\phi$-integral after multiplying by $\hat n_x$ or $\hat n_y$.  This implies that we need only consider the $p=\pm 1$ contributions to $F_x$ and $F_y$, and moreover that only resonances with $|n_\phi|=1$ can contribute.
Moreover, under reflection across the equator, $F_x$ and $F_y$ must remain unchanged but $\psi^\theta$ is incremented by $\pi$ (the ascending node becomes a descending node); therefore there can be no contribution to Eq.~(\ref{eq:FluxEm}) from terms with $pn_\theta$ odd.  Thus the only resonances that contribute to $F_x$ and $F_y$ have $|n_\phi|=1$ and $n_\theta$ even.
\end{itemize}

The resonance condition (Eq.~\ref{eq:rescond}) with $(n_\theta,n_\phi)$ is also satisfied for $(-n_\theta,-n_\phi)$; so in what follows we simply choose the positive sign for $n_\phi=+1$.

The resonances at which there is a net force on the binary are thus those where $\Omega_\phi:\Omega_\theta$ are in the ratios 2:1, 4:1, 6:1, etc., corresponding to $n_\theta=-2$, $-4$, $-6$, etc.  All of these resonances can in principle occur, but only for rapidly spinning primary holes, prograde orbits, and small radii.  We can see this by considering a maximally spinning black hole, $a_\star=1$, and small inclinations $\iota\ll 1$, for which \cite{1990PASJ...42...99K}
\begin{equation}
\frac{\Omega_\phi}{\Omega_\theta} = \left[ 1 - 4(r/M)^{-3/2} + 3(r/M)^{-2} \right]^{-1/2};
\end{equation}
this approaches $1$ at large radii and $\infty$ as $r\rightarrow r_{\rm ISCO}=M$.

In practice, evaluating the ratio $\Omega_\phi/\Omega_\theta$ at the ISCO for different $a_\star$ enables us to determine the minimum black hole spin at which each resonance is possible.  Thus we see that the 2:1 resonance can exist for $a_\star>0.9678$; the 4:1 resonance can exist for $a_\star>0.99722$; and the 6:1 resonance can exist for $a_\star>0.999253$.  We thus expect that the 2:1 resonance will be relevant over the largest region of parameter space.  We emphasize, however, that in order to obtain nonzero emitted amplitudes for $k\neq 0$ the inclination cannot be exactly zero.

\begin{figure*}
\includegraphics[angle=-90,width=6.5in]{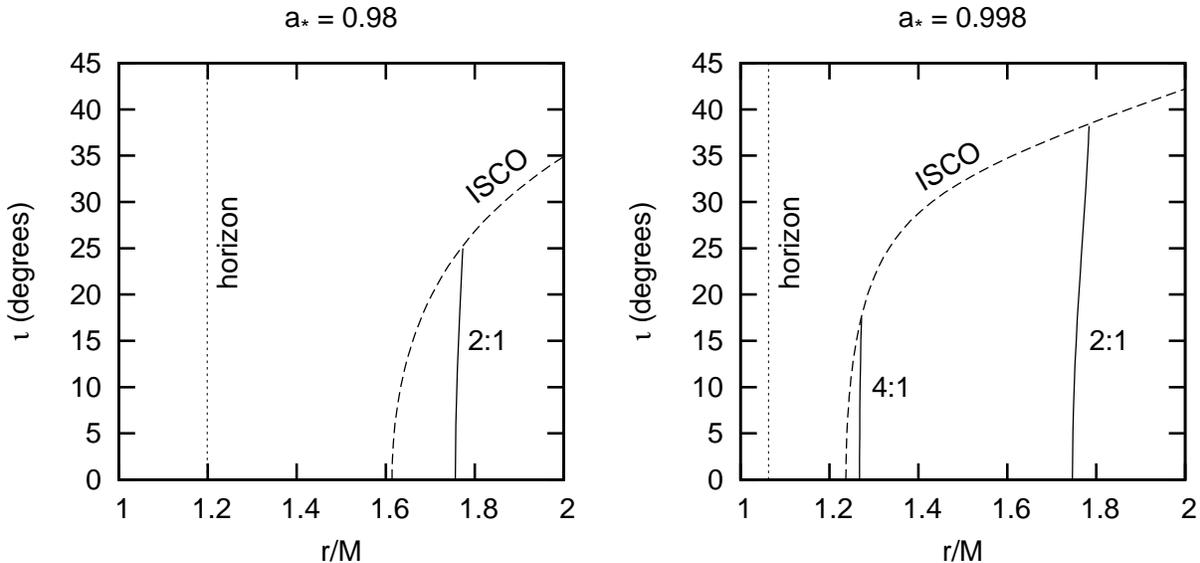}
\caption{\label{fig:resloc}The resonance locations in the $(r,\iota)$-plane for $a_\star = 0.98$ and 0.998.  The vertical dotted line on the left in each panel shows the horizon, and the dashed curve shows the ISCO location.  The solid curves show the resonance locations.  Additional resonances appear as we increase $a_\star$.}
\end{figure*}

The resonance locations for two values of $a_\star$ are displayed in Fig.~\ref{fig:resloc}.  The value $a_\star=0.998$ is a rough upper limit to the spin parameter of a black hole spun up by accretion due to capture of negative angular momentum radiation \cite{1974ApJ...191..507T}; $a_\star=0.98$ shows a less extreme case that still possesses a 2:1 resonance.  Note that as $a_\star$ increases, additional resonances appear.

\subsection{Kick magnitude}

In order to calculate the magnitude of the kick received at the resonance, we need to know how long the force ${\bf F}$ acts.  In the limit of an idealized test particle following a true geodesic of the Kerr metric, $\mu\rightarrow 0$, the mean force ${\bf F}$ acts for an infinite amount of time.  However at this point, it is necessary to consider radiation reaction effects.  In particular, the resonant argument
\begin{equation}
\varphi \equiv n_\theta \psi^\theta + \psi^\phi
\end{equation}
does not remain constant.  Rather, as the particle inspirals and approaches the resonance it circulates at some rate $\dot\varphi = n_\theta\Omega_\theta + \Omega_\phi < 0$.  (Remember that $n_\theta$ is a negative even integer.)  As it drifts inward, $\dot\varphi$ increases linearly with time and crosses zero at the resonance.  Thereafter it is positive and the resonant argument begins to circulate again.  The time over which the resonant argument can be treated as roughly constant is $\approx|\ddot\varphi|^{-1/2}\propto\mu^{-1/2}$, and there will be an overall recoil kick velocity ${\bf V} = -\int {\bf F}\,dt/M$ proportional to $\mu^{3/2}$.

The stationary phase approximation provides a more mathematical statement of the above argument, and is the method by which we may obtain the numerical prefactor.  Indeed, the reason that circular, equatorial inspirals into Kerr lead to only a $\mu^2$ contribution to the kick is precisely that there is no Fourier component of the emitted force \cite{2010PhRvD..81j4009S} that passes through a stationary phase point.

In order to determine the {\em magnitude} of the kick there is no need in this problem for a ``self-force'' calculation: the conservative part of the self-force should yield an ${\cal O}(\mu)$ shift in the resonance position, but it is the dissipative part (i.e. the part that changes the actions) that determines $\ddot\varphi$.  The self-force is required to obtain the direction of the kick from initial data at large separations because one needs to know the longitude of the ascending node at resonance crossing, and hence the phase evolution must be accurate to $<1$ radian.  In the case of inspirals that pass through multiple resonances (2:1 and then 4:1, and possibly higher) even the magnitude of the total kick depends on phase evolution because one must know the longitude of ascending node advance between successive resonance crossings.  These issues are beyond the capabilities of the code in \cite{2010arXiv1010.0759H} and hence will not be explored here.

We may now apply the stationary phase approximation to the $p=\pm 1$ contributions to Eq.~(\ref{eq:FluxEm}).  It is easily seen that the two contributions are complex conjugates of each other, so
\begin{eqnarray}
\left.\frac{dE}{dt\,d^2\hat{\bf n}}\right|_{p=\pm1} \!\!\! &=& \!\!\!
\frac{\mu^2}{4\pi}
\sum'_{\ell_1 m_1 k_1 \ell_2}
\omega^{-2} \tilde Z^{{\rm out}\ast}_{\ell_1 m_1k_1} \tilde Z^{{\rm out}}_{\ell_2 m_2k_2}
\nonumber \\ && \!\!\! \times
S_{\ell_1 m_1 \omega}(\theta) S_{\ell_2 m_2 \omega}(\theta)
e^{i(\phi - \varphi)}
\nonumber \\ && \!\!\!
+{\rm c.c.},
\end{eqnarray}
where $m_2=m_1+1$ and $k_2 = k_1 + n_\theta$.  Integrating the $x$ and $y$ components of the force using $\hat n_x - i\hat n_y = \sin\theta\,e^{-i\phi}$ gives a force:
\begin{eqnarray}
F_x - iF_y \!\!\! &=& \!\!\! \frac{\mu^2}{2} e^{ - i\varphi } \int_0^\pi
\sum'_{\ell_1 m_1 k_1 \ell_2}
\omega^{-2} \tilde Z^{{\rm out}\ast}_{\ell_1 m_1k_1} \tilde Z^{{\rm out}}_{\ell_2 m_2k_2}
\nonumber \\ && \!\!\! \times
S_{\ell_1 m_1 \omega}(\theta) S_{\ell_2 m_2 \omega}(\theta)
\,\sin^2\theta\,d\theta.
\label{eq:ftemp}
\end{eqnarray}
(The complex conjugate term has no contribution here; it contributes instead to $F_x+iF_y$.)

Now as we sweep over the resonance, the integral in Eq.~(\ref{eq:ftemp}) varies slowly; the evolution as one moves slightly off resonance is dominated by the $e^{-i\varphi}$ term.  If we Taylor-expand $\varphi$ as
\begin{equation}
\varphi = \varphi_0 + \frac12\ddot\varphi(t-t_0)^2 + ...
\end{equation}
then we have the time integral
\begin{equation}
\int_{-\infty}^\infty e^{-i\varphi}\,dt \approx \sqrt{\frac{2\pi}{|\ddot\varphi|}} \exp\left( -i\varphi_0 - i \frac\pi4\,{\rm sgn}\,\ddot\varphi \right).
\end{equation}
(This is the essence of the stationary phase approximation.)
Using this to integrate Eq.~(\ref{eq:ftemp}), and using $\ddot\varphi = A\mu$ with $A$ independent of $\mu$, we see that the kick velocity is
\begin{eqnarray}
|{\bf V}| \!\!\! &=& \!\!\!
\sqrt{\frac\pi{2|A|}}
\,\mu^{3/2}\Biggl|\,
\int_0^\pi
\sum'_{\ell_1 m_1 k_1 \ell_2}
\omega^{-2} \tilde Z^{{\rm out}\ast}_{\ell_1 m_1k_1} \tilde Z^{{\rm out}}_{\ell_2 m_2k_2}
\nonumber \\ && \!\!\! \times
S_{\ell_1 m_1 \omega}(\theta) S_{\ell_2 m_2 \omega}(\theta)
\,\sin^2\theta\,d\theta\, \Biggr|.
\label{eq:vres}
\end{eqnarray}

Equation~(\ref{eq:vres}) is the main theoretical result of this paper: it establishes the existence of a kick during the inspiral phase of order $\mu^{3/2}$ under any circumstance that leads to a resonant crossing.  Nevertheless, we must still compute the rescaled phase acceleration $A$ at the resonance.
It is
\begin{equation}
A = \sum_i \frac{\dot {\tilde J_i}}\mu \left( n_\theta \frac{\partial \Omega_\theta}{\partial \tilde J_i} + \frac{\partial\Omega_\phi}{\partial\tilde J_i} \right).
\label{eq:A}
\end{equation}
Since we consider circular orbits, $J_r=0$; in this case, the rate of energy and angular momentum loss are sufficient to follow the remaining two ``constants'' of the motion \cite{2000PhRvD..61h4004H, 2001PhRvD..64f4004H}.  Here the rate of change of angular momentum is (in the notation of Ref.~\cite{2010arXiv1010.0759H})
\begin{equation}
\dot{\cal L} = \dot{\tilde J_\phi} = -\mu \sum'_{\ell m k} \frac{m}{2\omega_{mk}^3} \left(
|\tilde Z^{\rm out}_{\ell mk}|^2
+ \alpha_{\ell mk}|\tilde Z^{\rm down}_{\ell mk}|^2
\right),
\label{eq:ldot}
\end{equation}
where $\alpha_{\ell mk}$ depends on the separation constants and $\tilde Z^{\rm out}_{\ell mk}$ and $\tilde Z^{\rm down}_{\ell mk}$ are the wave amplitudes emitted to future null infinity and into the future horizon (divided by $\mu$).  We can obtain $\dot{\cal E}$ by making the replacement $m/(2\omega_{mk}^3) \rightarrow 1/(2\omega_{mk}^2)$; and then we can obtain $\dot{\tilde J_\theta}$ using
the fact that the fundamental frequencies are the partial derivatives of the energy with respect to the actions, and hence
$\dot{\cal E} = \Omega_\theta \dot{\tilde J_\theta} + \Omega_\phi \dot{\tilde J_\phi}$.  Since $\omega_{mk} = k \Omega_\theta + m\Omega_\phi$, we can infer $\dot{\tilde J_\theta}$:
\begin{equation}
\dot{\tilde J_\theta} = -\mu \sum'_{\ell m k} \frac{k}{2\omega_{mk}^3} \left(
|\tilde Z^{\rm out}_{\ell mk}|^2
+ \alpha_{\ell mk}|\tilde Z^{\rm down}_{\ell mk}|^2
\right).
\label{eq:jtdot}
\end{equation}

\section{Results}
\label{S:r}

We are now in a position to evaluate the resonant kick contribution.  We focus on the 2:1 resonance, which occurs in the largest region of parameter space.  The radiated coefficients $\tilde Z^{\rm out}_{\ell mk}$ and $\tilde Z^{\rm down}_{\ell mk}$ are calculated here using the code of Ref.~\cite{2010arXiv1010.0759H}.  Inclinations reported here are defined using the Carter constant as in Ref.~\cite{2000PhRvD..61h4004H}: $\tan\iota \equiv \sqrt{\cal Q}/{\cal L}$.  The inclination of the initial orbit $\iota_{\rm init}$ is not exactly the same as the inclination $\iota_{\rm res}$ at resonance crossing, but the changes in inclination during the inspiral are small so we will not integrate the full trajectory through the $(r,\iota)$-plane in this paper \cite{2001PhRvD..64f4004H}.

As a specific example, we consider an inspiral into a primary black hole with spin $a_\star=0.998$ and inclination $\iota=20^\circ$.  The 2:1 resonance crossing is at $r=1.761M$, with actions $\tilde J_\theta = 0.09485M$ and $\tilde J_\phi = 1.599M$, energy ${\cal E}=0.7659$, Carter constant ${\cal Q}=0.3386M^2$, and frequencies $\Omega_r=0.0555M^{-1}$, $\Omega_\theta=0.1518M^{-1}$, and $\Omega_\phi=0.3037M^{-1}$.  There is no 4:1 or higher resonance crossing in this case because one reaches the last stable orbit ($\Omega_r=0$) first.

Considering modes with $\ell \le 12$, $|k|\le 6$, we derive a force at resonance of $|{\bf F}| = 4.15\times 10^{-4}\eta^2$ and an emitted power of $\dot E_{\rm em} = 6.74\times 10^{-2}\eta^2$.  (The gravitational-wave rocket is not very efficient: the asymmetry of the emitted power is only $|{\bf F}|/\dot E_{\rm em} = 0.006$.)  This leads to an inspiral rate of $\dot J_\theta = -0.00794\eta^2M$ and $\dot J_\phi = -0.197\eta^2M$, or $\dot r = -0.594\eta^2$ and $\dot\iota = 0.012\eta^2M^{-1}$.  (The inclination change is small compared to the radius change, in accordance with previous work \cite{2001PhRvD..64f4004H, 2002PhRvD..66f4005G}.)  The second derivative of the resonant argument is $\ddot\varphi = 0.129\eta M^{-2}$, leading to a total kick of ${\bf V} = 2.89\times 10^{-3}\eta^{3/2}$.  This appears to be converged with respect to the maximum $\ell$ and $|k|$: the kick increases by 2.4\%\ if we go to $\ell\le 16$, $|k|\le 8$, and by a further 0.4\%\ if we go to $\ell\le 20$, $|k|\le 10$.

\begin{figure}
\includegraphics[angle=-90,width=3.25in]{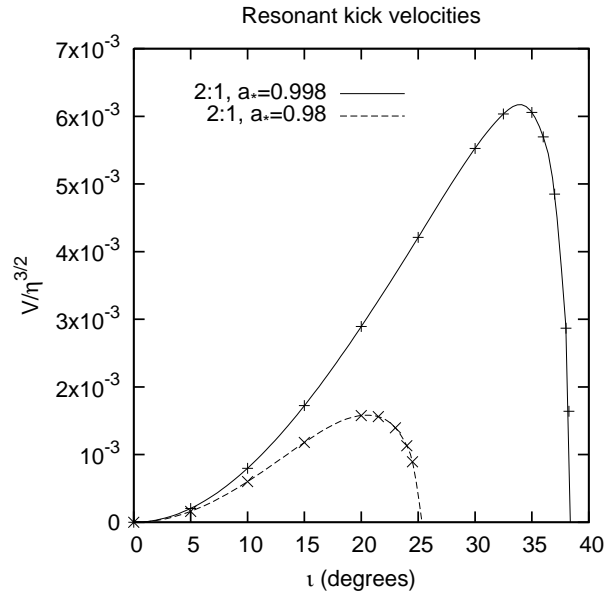}
\caption{\label{fig:pkick}The resonant kick velocity, with the factor of $\eta^{3/2}$ divided out.  We show curves including modes with $\ell\le12$ and $|k|\le 6$.}
\end{figure}

We have explored the behavior of the resonant kick velocity as a function of the primary spin and inclination.  The results are shown in Figure~\ref{fig:pkick} as a function of inclination for $a_\star=0.98$ and 0.998.  We can see the expected qualitative result that the kick velocity vanishes for equatorial orbits (it is $\propto\iota^2$ for small $\iota$).  As the inclination increases, the resonant kick also increases, until one approaches the maximum inclination at which the resonance lies outside the ISCO (this is 38$^\circ$ for the 2:1 resonance and $a_\star=0.998$).  There the inspiral velocity $|\dot r|$ becomes very large and $A\rightarrow\infty$.  As a consequence, the resonant kick, which according to the SPA is proportional to $|A|^{-1/2}$, drops to zero at the maximum inclination.

The analysis we have done here assumes that the dephasing time -- i.e. the time over which ${\bf F}$ deviates by $\pi/2$ from its the stationary direction, $t_{\rm d} = \pi^{1/2}|\ddot\varphi|^{-1/2}$ -- is much longer than the period of vertical oscillations, $T_\theta$; using $\ddot\varphi = A\mu$, we see that this condition is equivalent to
\begin{equation}
\eta \ll \eta_{\rm cr} \equiv \frac{4\pi}{MAT_\theta^2}.
\label{eq:etacr}
\end{equation}
(Since $A\propto M^{-3}$ and $T_\theta\propto M$ if we change $M$ while keeping the dimensionless parameters $a_\star$ and $\iota$ fixed, it follows that this condition on $\eta$ does not depend on $M$.)  For the above example of the 2:1 resonance with $a_\star=0.998$ and $\iota=20^\circ$, we find $\eta_{\rm cr} = 0.014$.  For mass ratios of order $\eta\sim \eta_{\rm cr}$, computing the resonant kick will require a more sophisticated analysis, likely one that actually follows the inspiral trajectory rather than using the stationary phase approximation.

The kick velocities associated with the 4:1 resonance are much smaller: for $a_\star=0.998$, we find a maximum resonant kick velocity of $\sim 3\times 10^{-5}\eta^{3/2}$ at inclinations $\iota\sim15^\circ$.  While this still becomes larger than the plunge kick ($\propto\eta^2$) in the extreme mass ratio limit $\eta\rightarrow 0^+$, the 4:1 resonant kick velocities are two orders of magnitude less than what we obtain from the 2:1 resonance and thus are subdominant.

\section{Discussion}
\label{S:d}

We have shown that the passage of an extreme mass ratio binary black hole through a resonance can lead to a fractional-order kick $V\propto \eta^{3/2}$ that dominates over the usual nonresonant inspiral + transition + plunge + ringdown kick ($\propto\eta^2$) for small $\eta$.  We exhibited this effect for the case of inclined circular inspirals into a rapidly rotating primary black hole, since the decay of eccentricity during the weak-field phase of the inspiral makes the circular orbit the most interesting case.  However, the resonant kick phenomenon should also occur for other types of EMRIs, including eccentric inspirals into Schwarzschild black holes, eccentric equatorial inspirals into Kerr holes, and generic (eccentric and inclined) inspirals into Kerr.  The formalism to describe these cases would be similar to that used here, but actual computations are left to future work.

The resonant kick described here is only one possible resonant interaction in extreme mass ratio binary black hole systems.  Flanagan \& Hinderer \cite{2010arXiv1009.4923F} have considered resonances between the vertical and radial motions in generic orbits, i.e. where $n_r\Omega_r+n_\theta\Omega_\theta=0$ for $n_r,n_\theta\in{\mathbb Z}$.  Unlike the resonances considered here, the phase at which one passes through a vertical-radial resonances actually has an influence on the change in the constants of motion $({\cal E}, {\cal Q},{\cal L})$, and hence leads to large ($\propto\eta^{-1/2}$) changes in the subsequent phase evolution of the inspiral, which would be significant for template construction.  The resonant recoil effect considered in this paper does not lead to such an intrinsic phase shift.  It does lead to a Doppler shift of the template before versus after the resonance passage, but since the fractional frequency shift is $\propto\eta^{3/2}$ and the number of cycles between resonance passage and plunge is $\propto\eta^{-1}$, the overall phase shift is $\propto\eta^{1/2}$ and hence will be small in the EMRI limit.

It seems unlikely (although not impossible) that the resonant kick is directly relevant in any astrophysical applications.  In the example of Section~\ref{S:r} of primary spin $a_\star=0.998$ and inclination $\iota=20^\circ$, the kick is $1.4(\eta/\eta_{\rm cr})^{3/2}\,$km$\,$s$^{-1}$ with $\eta_{\rm cr}=0.014$ being the maximum mass ratio for which we expect our analysis to hold (Eq.~\ref{eq:etacr}).  If $\eta\ll\eta_{\rm cr}$ this is small compared to the velocity dispersion of any system that could conceivably host a binary black hole.  Depending on the actual behavior of the resonant kick when $\eta/\eta_{\rm cr}$ is of order unity, it {\em might} be significant for globular clusters in the case of a stellar mass and intermediate mass black hole binary.  Even then one would have to arrange for a very large primary hole spin and the appropriate geometry.

Nevertheless, the existence of the new $\eta^{3/2}$ kick contribution suggests that the existing kick fitting formulae be treated with caution, particularly in the large-to-extreme mass ratio limit.  It also highlights the importance of exploring the $\eta \ll 0.1$ regime with numerical GR simulations \cite{2009PhRvD..79l4006G, 2010PhRvL.104u1101L, 2010arXiv1008.4360L, 2010arXiv1009.0292L}; this may reveal the behavior of the resonant kick at $\eta\sim\eta_{\rm cr}$ and elucidate the transition from $\eta^2$ to $\eta^{3/2}$ scaling.

\section*{Acknowledgements}

C.H. thanks Yanbei Chen, Tanja Hinderer, and Michael Kesden for helpful conversations and encouragement.

C.H. is supported by the US National Science Foundation (AST-0807337), the US Department of Energy (DE-FG03-02-ER40701), the Alfred P. Sloan Foundation, and the David and Lucile Packard Foundation.

\bibliography{kick}

\end{document}